\documentclass{icrc29}

\usepackage[mathscr]{eucal}
\usepackage{amsmath}
\usepackage{amssymb,amsfonts}
\usepackage{eucal}
\usepackage{subfigure}
\usepackage{epsfig}

\begin{document}

\newcommand*{\fig}{{Fig.}}

\title{{Search for localized excess fluxes in Auger sky maps and
    prescription results}}

\presenter{Pierre Auger Collaboration\\
{\it Pierre Auger Observatory, av. San Mart\'{\i}n Norte 304, (5613)
  Malarg\"ue, Argentina}\\
Presenter: B. Revenu (revenu@iap.fr), fra-revenu-B-abs1-he14-oral}

\maketitle

\begin{abstract}
  Using the first surface detector data of the Pierre Auger
  Observatory, we present the results of a blind search for
  overdensities in the cosmic ray flux with respect to iso\-tro\-pic
  expectations.  We consider two energy bands: $1~\mathrm{EeV}
  \leqslant E \leqslant 5$ EeV and $E \geqslant 5$ EeV at two angular
  scales: 5$^\circ$ and 15$^\circ$. We also report the results of
  searches for excesses in target directions already defined in a set
  of prescriptions presented at the ICRC in 2003.  Both analyses give
  results that are compatible with isotropy.
\end{abstract}

\section{Introduction}

Since the beginning of the year~2004, the Pierre Auger Observatory has
achieved a long period of stable data acquisition which has allowed
systematic data analysis.  One of the main scientific goals of the
experiment is the study of the arrival directions of cosmic rays.
Statistically significant deviations from isotropy would give valuable
information on cosmic rays sources.  The AGASA~\cite{agasa} and
SUGAR~\cite{sugar} experiments have reported excesses in directions
close to the galactic center within a limited energy range.  In this
article, we will focus on a blind search over the whole sky and on
targeted searches as defined in the prescription presented at the ICRC
in~2003~(see~\cite{clay2003}).  To detect an excess of events coming
from any particular region of the sky, we have to compare the observed
number of events with that expected from an isotropic flux of cosmic
rays, taking into account the relative exposure of the different sky
directions. The significance of the resulting excess and deficit map
has to be determined and the distribution of significances has to be
compared with that obtained for a large number of isotropic
simulations.

\section{The dataset}\label{dataset}

We are considering the surface detector data only, without making use of
fluorescence data and without using
a constant intensity cut as it is done in the spectrum paper
(see~\cite{spectre}) and in the Galactic center studies paper
(see~\cite{AugerGC}).

In the prescription, the data period is defined as the time interval
between 8 August 2003 to 16 May 2004. In this work, because of some
trigger instability, we use actually data from 1 January 2004 to 16
May 2004. This represents~6046 events with zenith angles within
60$^\circ$ of vertical and that have passed all quality reconstruction
cuts.  The angular accuracy is always smaller than 2.2$^\circ$
(see~\cite{angularICRC}) allowing us to search for fluctuations on
scales of a few degrees.  For the blind search, we use the data from 1
January 2004 to 11 May 2005 with the same cuts on zenith angle and
quality criteria. We have $30548$ events, $29073$ in the range
${1~\mathrm{EeV}\leqslant E \leqslant 5}$~EeV and $1475$ above 5~EeV.
The energy of 5~EeV corresponds approximately to the ankle in the
cosmic ray spectrum, which may represent the transition from a
galactic to extragalactic component.  We choose to explore the data in
angular scales of 5$^\circ$ and 15$^\circ$ radius since excess events
from point sources may have arrival directions spread by intervening
magnetic fields and because the statistics is still not that large to
justify a blind search at the scale of our angular resolution.

\section{Methods}\label{methods}

\subsection{Coverage map}

The coverage map gives the relative number of expected events arriving
to the detector from any direction in the sky for an isotropic
incoming CR flux. Various methods exist to estimate it, such as the
shuffling method consisting of exchanging arrival ti\-mes and azimuths
of real events to generate fake data sets that are averaged on the sky
to obtain the coverage map. There is the semi-analytic one based on a
smooth fit of the zenith angle distribution of the real events. The
semi-analytic method takes into account the geometrical acceptance
modulation, the zenith angle cutoff and the right-ascension modulation
due to the non uniformity of the surface detector deployment and
running. These methods are discussed in~\cite{coverageICRC}. Note that
for small angular scales both methods lead to comparable results since
the error induced is always smaller than the Poisson noise of the
event sample.

\begin{center}
  \begin{figure}[!ht]     
      \begin{center}
        \includegraphics[draft=false,scale=0.29,angle=90]{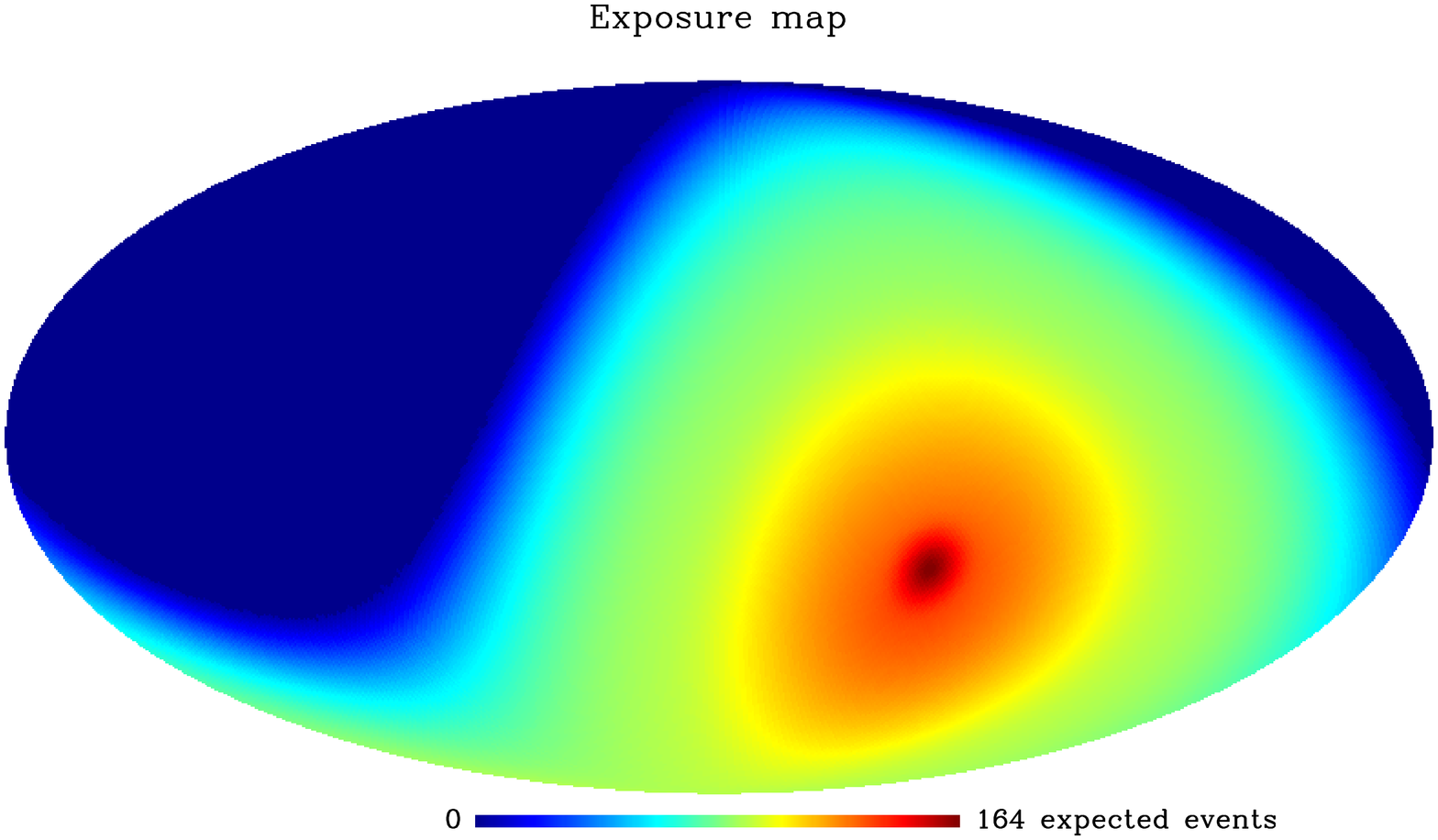}
        \includegraphics[draft=false,scale=0.29,angle=90]{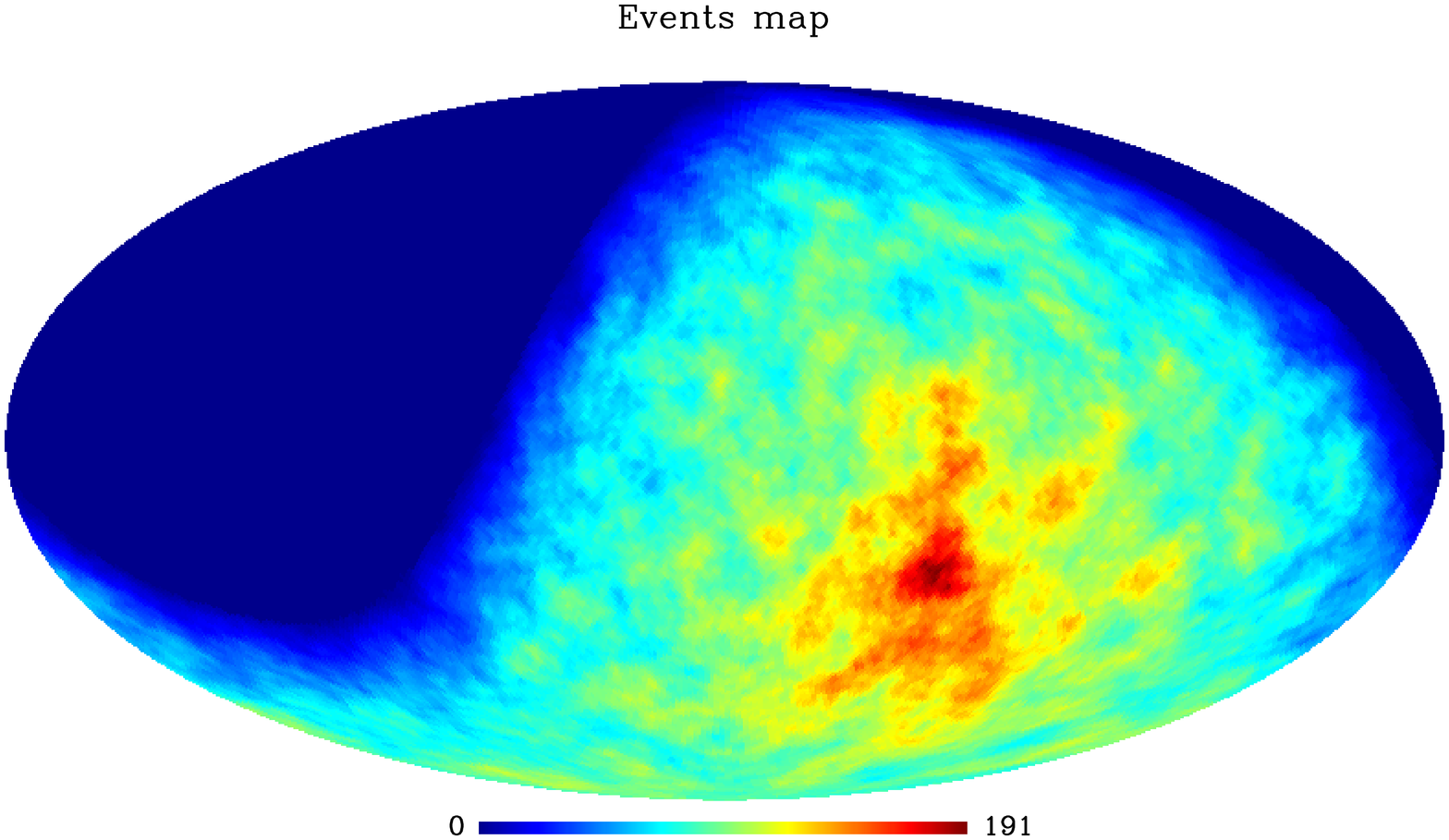}
        \caption{Left: semi-analytic Auger south sky coverage in the
          energy range [1-5]~EeV. Right: real event map. The two sky maps
          are in galactic coordinates.}\label{coveragefig}
      \end{center}
  \end{figure}
\end{center}

In \fig~\ref{coveragefig} we present the sky coverage map (left) and
the actual events map (right) for the energy range [1-5]~EeV smoothed
with a top-hat \footnote{Selects data satisfying $\arccos
  (\hat{\mathbf{n}}_\mathbf{w} \cdot
  \hat{\mathbf{n}}_\mathbf{d})\leqslant \alpha$ where
  $\hat{\mathbf{n}}_\mathbf{w}$ is the direction of the center of the
  window and $\hat{\mathbf{n}}_\mathbf{d}$ is the direction of the
  data and $\alpha$ is the radius of the window.} beam of radius
5$^\circ$. We use the Healpix~\cite{gorski98} pixellisation of the
sphere.

The coverage map depends on the energy range. The appropriate map is
used for each particular search, either blind search for a particular
energy range or prescription search.

\subsection{The significance map}\label{statistics}

For each pixel of the sky, we compute the number of expected events
integrating the semi-analytical coverage map in a given window
(top-hat, gaussian...) centered on this pixel.  Then the signal is
obtained by applying the same filtering but on the event map.  The
significance of the signal with respect to the expected background is
computed following~\cite{lima} in the case of a top-hat window and
will be referred hereafter by the expression~: ''Li-Ma significance''.

\section{Blind search for overdensities}\label{blindsearch}

\begin{figure}[!ht]     
  \begin{center}
    \includegraphics[draft=false,scale=0.3]{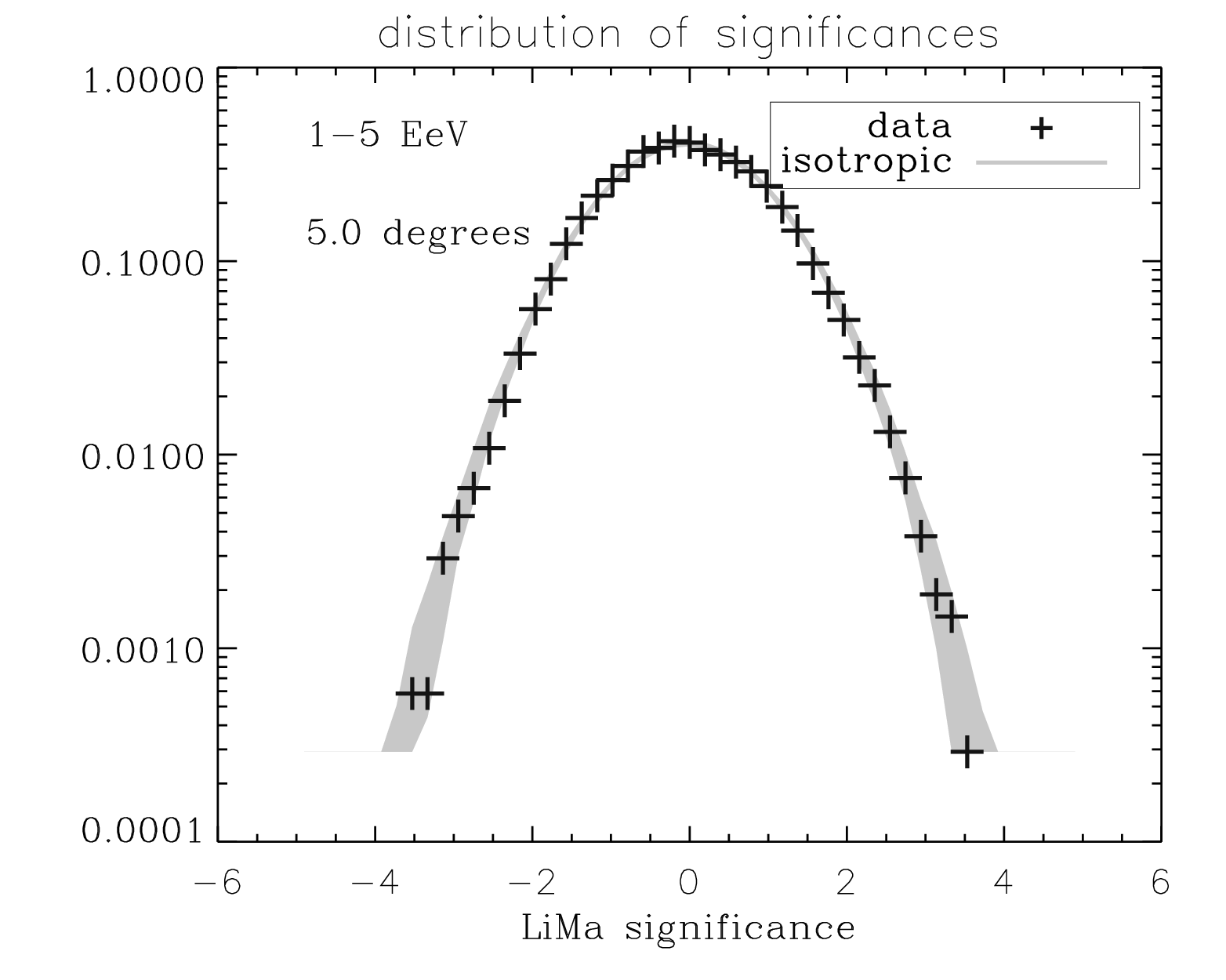}
    \includegraphics[draft=false,scale=0.3]{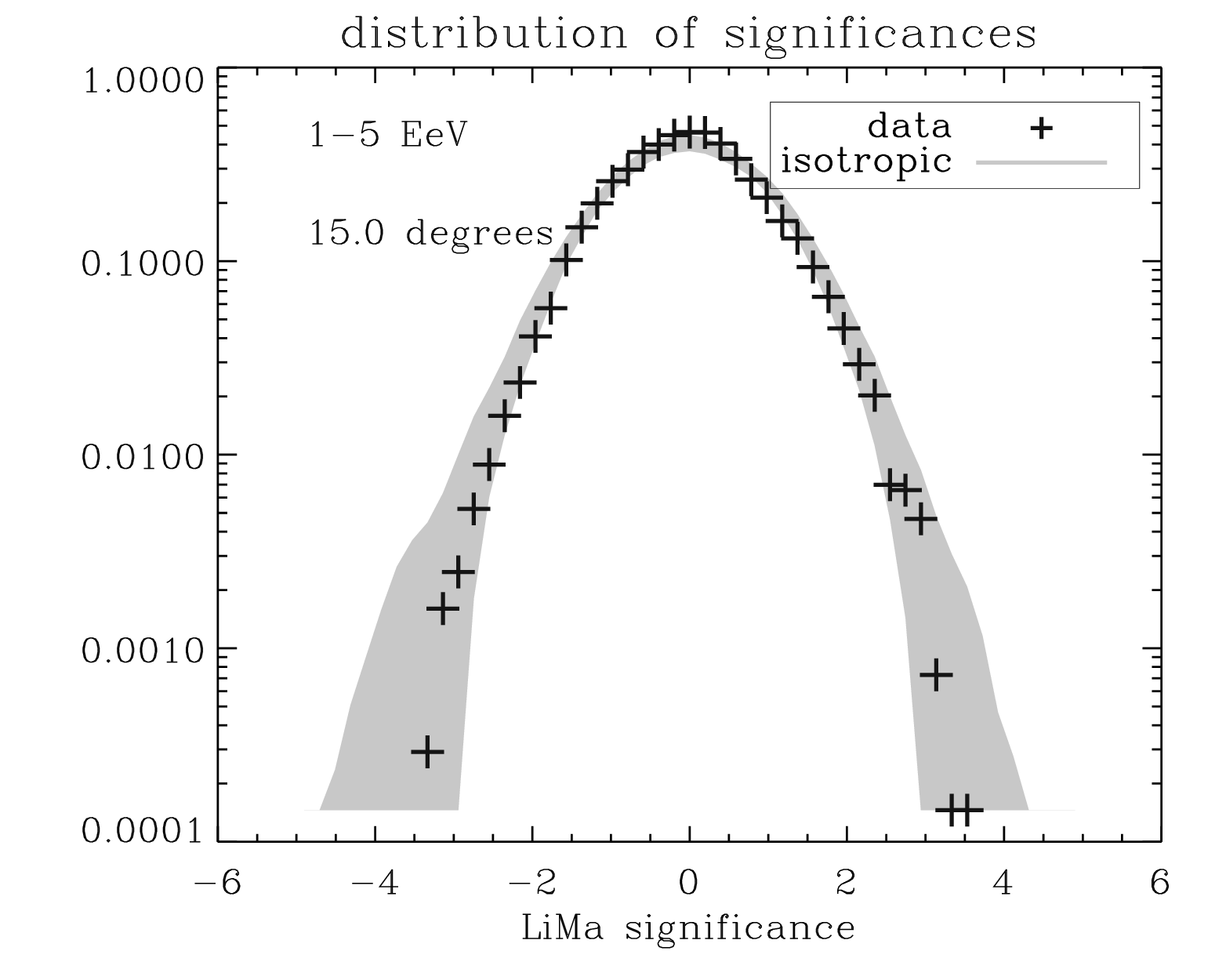}
    \includegraphics[draft=false,scale=0.3]{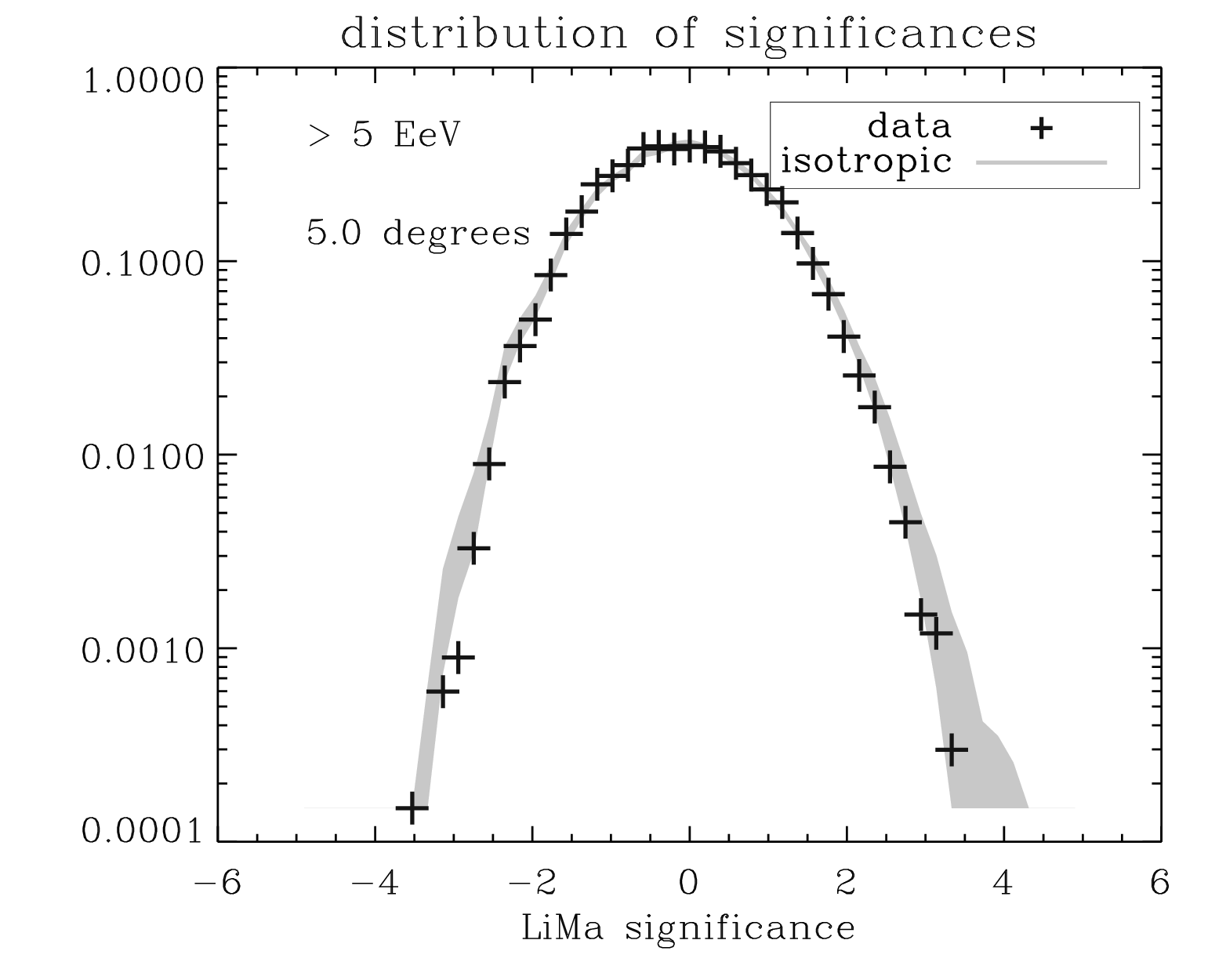}
    \includegraphics[draft=false,scale=0.3]{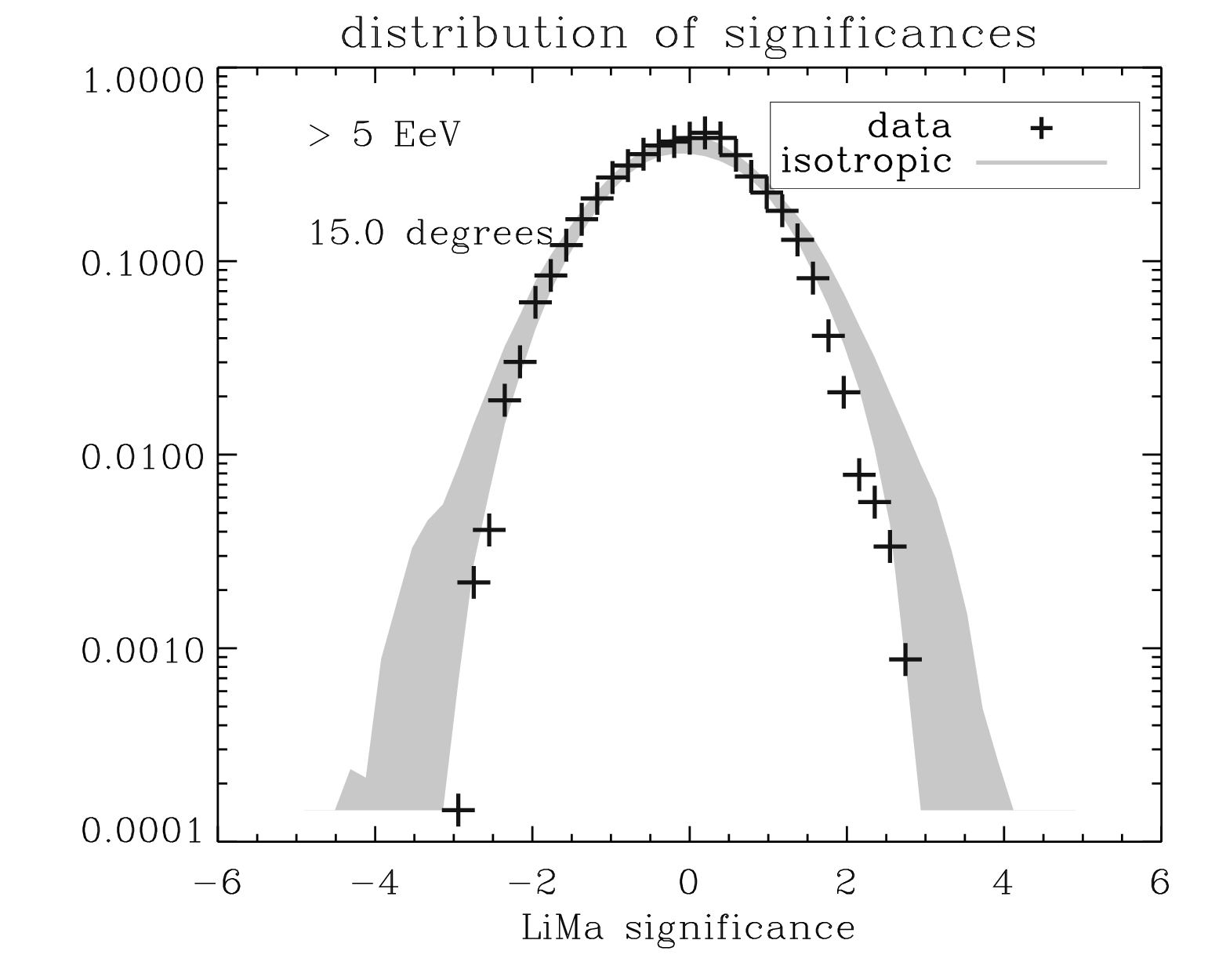}
    \caption{Li-Ma significance distributions for angular scales of
      5$^\circ$ and 15$^\circ$ for $1~\mathrm{EeV}\leqslant E \leqslant 5$~EeV and
      $E \geqslant 5$~EeV. The number of density fluctuations at a
      given significance level is compatible with what we expect from
      isotropic simulated sky (shaded region).}\label{limadistribs1}
  \end{center}
\end{figure}

In order to see if the largest overdensities found are compatible with
those expected from an isotropic CR flux, we compare the distribution
of the significances obtained with those obtained for a large number
of isotropic Monte~Carlo simulations. These distributions are shown in
\fig~\ref{limadistribs1}.  The shaded areas correspond to the
1~$\sigma$ dispersion of the isotropic simulations.  In all the cases
the distribution is consistent with isotropic expectations.

\section{Prescription results}\label{prescription}

At the 2003 ICRC, the Pierre Auger Collaboration presented a
prescription in order to properly compute the probability for a local
anisotropy to be compatible with isotropic expectations. Please refer
to~\cite{clay2003} for the detail.  This procedure avoids
uncertainties coming from the ``penalty factors'' due to {\bf a
  posteriori} probability estimations, which can overestimate the
significance of a source candidate.

The prescription targets are the galactic center and the AGASA\--SUGAR
location for the low energy data. For the highest energy events,
targets are three nearby violent extragalactic objects: NGC0253,
NGC3256 and Centaurus~A.

The properties of the prescription targets and the
results are shown in Table~\ref{prescriptions_results}.
\vspace{-0.5cm}
\begin{center}
  \begin{table*}[!htb]
    \begin{center}
      \begin{tabular}{|c|c|c|c|c|c|c|c|c|c|}
        \hline
        Target & $\ell (^\circ)$ & $b (^\circ)$ & Radius & $\log(E/\mathrm{EeV})$&
        Found & Exp. & Prob & Req. Prob\\ \hline
        GC 1 &  0.00 &  0.00 & $15^{\circ}$ &  $\geqslant 18$ & 155  &  167.3  & - & 0.0035\\
        GC 2 &  0.00 &  0.00 & Point ($2^{\circ}$)& $18-18.5$  & 2    &   2.5 & - & 0.00025\\
        AGASA SUGAR & 7.00 & 0.00 & Point ($2^{\circ}$)& $18-18.5$    &
        3   &   2.69 & 0.43 & 0.00025\\
         NGC0253  & 88.92 & -87.80&  $5^{\circ}$ & $\geqslant 19.5$    & 0    &     0.01   & - & 0.00005\\
        NGC3256 & 277.56  & 11.49& $5^{\circ}$ & $\geqslant 19.5$
        & 0    &     0.01   & - & 0.00005\\
         Centaurus A & 309.43  & 19.44 & $5^{\circ}$ & $\geqslant 19.5$        &  0   &  0.01  & - & 0.00005\\      
        \hline
      \end{tabular}
      \caption{Results for the ICRC 2003 prescription with the dataset defined in
        section~\ref{dataset}. Columns are: target name, galactic
        longitude, galactic latitude, radius of the top-hat window,
        energy range, observed number of events, expected number of
        events, probability associated to the Li-Ma significance,
        required probability level for positive detection. The
        probability is not computed in case of a deficit of events
        with respect to the background
        expectations.}\label{prescriptions_results}
    \end{center}
  \end{table*}
\end{center}
\vspace{-0.5cm} It can be seen that none of the prescription targets
leads to a positive detection.  No excess is found from the Galactic
Center region (see~\cite{AugerGC} for a complete analysis).

\section{Conclusion}\label{conclusion}

We have presented searches for localized excesses in the first Auger
sky maps, with an unprecedented accumulated statistics at ultra-high
energies in the Southern hemisphere. Blind source searches did not
reveal any remarkable excess in any of the two energy bands
$1~\mathrm{EeV}\leqslant E \leqslant 5$ EeV and $E \geqslant 5$ EeV.
We presented the results of the prescription which had been set in
2003 for the first Auger dataset. None of the prescribed targets leads
to a significant excess.  Studies are underway on larger datasets.

\end{document}